# Mask-less Patterning of Gallium-irradiated Superconducting Silicon Using Focused Ion Beam

Ryo Matsumoto[a,b], Shintaro Adachi[a], El Hadi S. Sadki[c], Sayaka Yamamoto[a,b], Hiromi Tanaka[d], Hiroyuki Takeya[a], and Yoshihiko Takano[a,b]

[a]National Institute for Materials Science, 1-2-1 Sengen, Tsukuba, Ibaraki 305-0047, Japan
[b]Graduate School of Pure and Applied Sciences, University of Tsukuba, 1-1-1 Tennodai, Tsukuba, Ibaraki 305-8577, Japan
[c]Physics Department, College of Science, United Arab Emirates University, Al Ain UAE
[d]National Institute of Technology, Yonago College, 4448 Hikona, Yonago, Tottori 683-8502, Japan

**Abstract**

A direct patterning technique of gallium-irradiated superconducting silicon has been established by focused gallium-ion beam without any mask-based lithography process. The electrical transport measurements for line and square shaped patterns of gallium-irradiated silicon were carried out under self-field and magnetic field up to 7 T. Sharp superconducting transitions were observed in both patterns at temperature of 7 K. The line pattern exhibited a signature of higher onset temperature above 10 K. A critical dose amount to obtain the superconducting gallium-irradiated silicon was investigated by the fabrication of various samples with different doses. This technique can be used as a simple fabrication method for superconducting device.

## 1. Introduction

It has been recently discovered that heavily hole-doped group-IV semiconductors like diamond [1-3], silicon [4], and germanium [5] exhibit superconductivity. According to a McMillan relation, superconducting transition temperature $T_c$ is proportional to the Debye temperature $θ$ [6]. The group-IV semiconductors, especially, diamond and silicon are candidates for high-$T_c$ superconductors because they show remarkable high Debye temperature [7]. Various methods to induce the high hole-carrier concentration in these materials have been studied. For example, high-pressure synthesis [8], chemical vapor deposition [9], electric field effects [10-12], and so on [13]. Ion implantation is a strong tool to induce hole carrier in these materials [14-16]. Recently, an appearance of superconductivity with $T_c$ at around 7 K was reported in gallium-doped silicon via the ion implantation process [17-20]. Although the discovery with relatively high $T_c$ above liquid helium temperature in silicon attracts wide attention, it is necessary to use a resist-based conventional lithographic process to obtain a desired pattern of a superconducting circuit. Such a patterning is crucial for the application of superconducting devices, for example, the superconducting quantum interference device (SQUID) magnetometer.

The focused ion beam (FIB) is one of the most popular nanofabrication techniques for semiconducting devices [21], ultra-thin films [22], and superconducting Josephson devises [23,24]. FIB nanofabrication provides direct etching and deposition for desired shapes without resist-based process. In general, the desired region of sample can be milled by scanning the ion-beam over it in the FIB fabrication. The etching rate is determined by the dose amount of gallium ions which are tuned by the condenser lens, aperture size and dose time. If the dose amount of FIB beam achieves a certain criterion for the emergence of superconductivity in silicon, it will be possible to obtain desired patterns of superconducting silicon directly without resist-based lithographic process.

In this study, we investigated the irradiation effect of accelerated gallium-ions on silicon substrate by FIB. The surface states of irradiated region were analyzed by a core-level X-ray photoelectron spectroscopy (XPS). Two kinds of line and square shaped gallium-irradiated silicon with different dose amount were evaluated using an electrical resistance measurement.

## 2. Experimental procedures

The gallium irradiations for silicon substrate were carried out using a SMI9800SE FIB machine (Hitachi High-Technologies), equipped with Ga$^+$ beam. The acceleration voltage of the ion beam, FIB current, chamber pressure were 30 keV, 3.6 μA with an aperture size of 2 mm, and $3×10^{-5}$ Pa, respectively. The dose amount of gallium ions, which is determined from the formula $IT/S$ where $I$ is FIB current, $T$ is dose time, and $S$ is dose area, was mainly adjusted by the dose time.

The surface states of gallium-irradiated region on silicon was analyzed by the core-level XPS (AXIS-ULTRA DLD, Shimadzu/Kratos) with AlK$α$ X-ray radiation ($hν$ = 1486.6 eV), operating under a pressure of the order of $10^{-9}$ Torr. The background signals were subtracted by using an active Shirley method on COMPRO software [25]. The photoelectron peaks were analyzed by the pseudo-Voigt functions peak fitting.

The gallium-irradiated region on silicon was evaluated from the temperature dependence of resistance via a standard four probe method using physical property measurement system (PPMS,

Quantum Design) with 7 T superconducting magnet. The electrodes were made by a silver-paste painting and gold wires on the irradiated region. The temperature dependence of upper critical field ($H_{c2}$) of the silicon was determined from onset $T_c$ value.

## 3. Results and discussion

Figure 1 (a) shows a temperature dependence of resistance in the gallium-irradiated silicon of line pattern (1×1000×2 μm) with dose amount of 5.7×10$^{18}$ C/cm$^2$. The resistance sharply dropped to zero with superconducting onset temperature ($T_c^{onset}$) of 7 K and zero resistance temperature $T_c^{zero}$ of 6 K in agreement with the previous report of the gallium-doped silicon induced by the ion-implantation and a rapid thermal treatment [18]. For the practical applications, the anisotropy of superconductivity was evaluated through a measurement for an angle dependence of the resistance under magnetic fields at 5 K as shown in Fig. 1 (b). The superconductivity was sensitively suppressed around 0 degree, namely under magnetic field perpendicular to the substrate. In contrast, the superconductivity was robust against the magnetic field parallel to the substrate. The temperature dependence of resistance was investigated under various magnetic fields (c) parallel and (d) perpendicular to the substrate. The insets show temperature dependence of $H_{c2}$. The critical fields follow a typical parabolic behavior, which is consistent with the previous report for the gallium-doped silicon [18]. Here, a maximum critical fields at zero temperature $H_{c2//}(0)$ and $H_{c2\perp}(0)$ corresponding to $H_{c2}(0)$ under a magnetic field which is parallel to the substrate and perpendicular to the substrate, respectively. The $H_{c2//}(0)$ of 14.8 T and $H_{c2\perp}(0)$ of 10.7 T were estimated by the parabolic fit. The anisotropic parameter $\gamma = H_{c2//}(0) / H_{c2\perp}(0)$ was 1.4. The coherence length at zero temperature $\xi_{//}(0)$ and $\xi_\perp(0)$ were estimated as 4.7 nm and 5.5 nm, respectively, from the Ginzburg–Landau (GL) formula $H_{c2}(0) = \Phi_0/2\pi\xi(0)^2$, where the $\Phi_0$ is the flux quantum.

To confirm a flexibility for fabrication of superconducting silicon, we also prepared a square pattern (200×200×0.3 μm) with dose amount of 8.5×10$^{17}$ C/cm$^2$ and evaluated in Fig. 1 (e-h) using same method with that in the line pattern. The square pattern exhibited almost same $T_c$ with that of line pattern. Slightly small $T_c^{onset}$ is maybe caused by lower dose amount. On the other hand, more emphasized antistrophic properties were observed as shown in Fig. 1 (f). The $H_{c2//}(0)$ of 18.8 T and $H_{c2\perp}(0)$ of 2.4 T were estimated by the parabolic fit. The anisotropic parameter $\gamma = H_{c2//}(0) / H_{c2\perp}(0)$ was 7.8, which is worthily higher than that of the line pattern, maybe reflecting its thin-film properties [18]. The coherence length at zero temperature $\xi_{//}(0)$ and $\xi_\perp(0)$ were estimated as 4.2 nm and 11.7 nm, respectively.

There are two possibilities for the origin of the superconductivity that is the aforementioned gallium-doped silicon and elemental $\beta$-gallium [26-28]. Here, the previously reported the critical field and the coherence length in the gallium-doped silicon and β-gallium are compared to those parallel to the substrate in gallium-irradiated silicon. According to the literatures [18, 28], the critical field and the coherence length were 9.4 T and 6 nm in the gallium-doped silicon, 57 mT and 76 nm in the β-gallium, respectively. Since the superconducting parameters from our product is similar to those from former, we conclude that the observed superconductivity in this study is originated from the gallium-doped silicon.

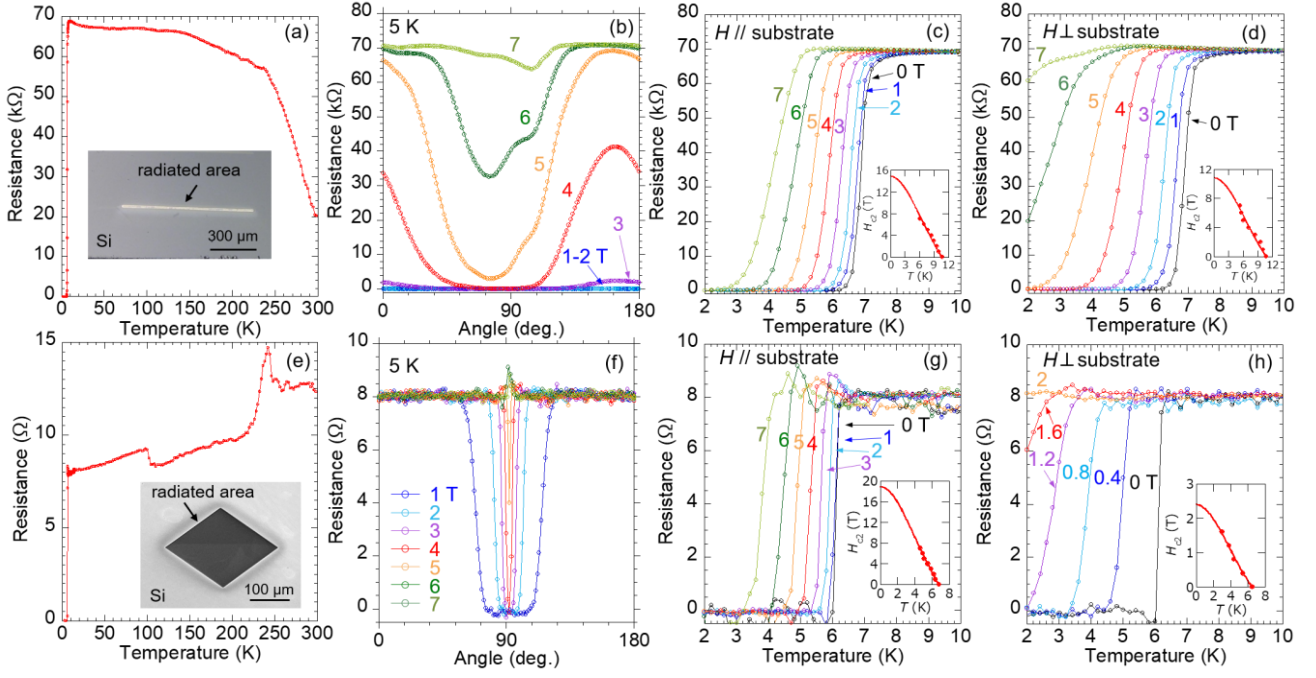

**Figure 1.** Superconducting properties in the gallium irradiated silicon of (a-d) line pattern (1×1000×2 µm) with dose amount of 5.7×10$^{18}$ C/cm$^2$ and (e-h) square pattern (200×200×0.3 µm) with dose amount of 8.5×10$^{17}$ C/cm$^2$. (a,e) Temperature dependence of resistance from 300 K to 2 K in square pattern. (b,f) Angle dependence of resistance at 5 K under various magnetic fields, (c,g) Enlargement around superconducting transition under various magnetic fields parallel to the line (or plane) and (d,h) perpendicular to the line (or plane).

Here, we note a specific temperature dependence of resistance under magnetic field in the gallium-irradiated silicon similar to a high-quality boron-doped superconducting diamond [29]. Figure 2 shows an enlargement of the temperature dependence of resistance in the gallium irradiated silicon of line pattern under magnetic fields. The separation between the resistances above 10 K and estimated the onset of transition to be at a value of ~12 K. The onset is gradually shifted to lower temperatures with increasing magnetic field. To clear the onset of transition, differential curves of resistance for temperature d$R$/d$T$ under 0 T and 7 T were shown in the inset of Fig. 3. We can see a clear separation of differential curves around 12 K under 0 T and 7 T. The signature of higher $T_c$ maybe attributed only to better crystallinity or a combination of better crystallinity and partially larger carrier density [29]. We can expect a bulk superconductivity above 10 K if the dose condition of gallium ion beam is optimized.

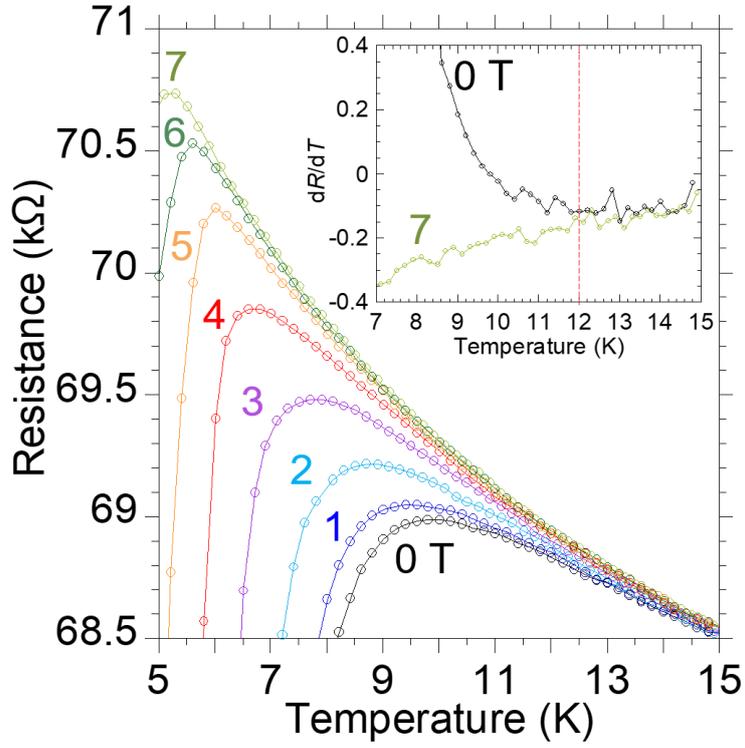

**Figure 2. Enlargement of temperature dependence of resistance in the gallium irradiated silicon of line pattern under magnetic fields. The inset is differential curve of resistance for temperature around $T_c^{onset}$.**

To confirm a substrate dependence of gallium-irradiation effect, we fabricated line-shaped pattern on various substrates with same dose amount of $5.7 \times 10^{18}$ C/cm$^2$ and measured their resistance-temperature (*R-T*) properties as shown in Fig. 3. The pattern dimensions were 1000 μm in length, 1 μm in width, 2 μm in depth, and the irradiation time is 1 hour. The gallium-irradiated silicon exhibited a semiconducting-like behavior and a sudden drop of resistance corresponding to superconductivity. To clarify the origin of superconductivity, we investigated the gallium-irradiation effects for various substrates. Although the gallium irradiated diamond substrate showed lower resistance than that of general undoped diamond, the drop of resistance was not observed in the *R-T* measurement. The gallium-irradiation effects for the conductive boron-doped diamond substrate and ITO (indium tin oxide) glass substrate were also investigated to exclude a charge-up effect during the gallium radiation as seen in high resistance substrates such an undoped diamond. As a result of the *R-T* measurements, both conductive substrates showed no superconductivity. The irradiated region on the insulating substrate of SiO$_2$ glass exhibited quite high resistance above 40 MΩ. In conclusion, the gallium-irradiated silicon is the only substrate that showed superconductivity, indicating that the origin of the superconductivity could be considered as a gallium-doped silicon as reported in the literature [17-20]. If superconducting gallium is deposited on the surface of the substrates, all substrates should show the superconductivity.

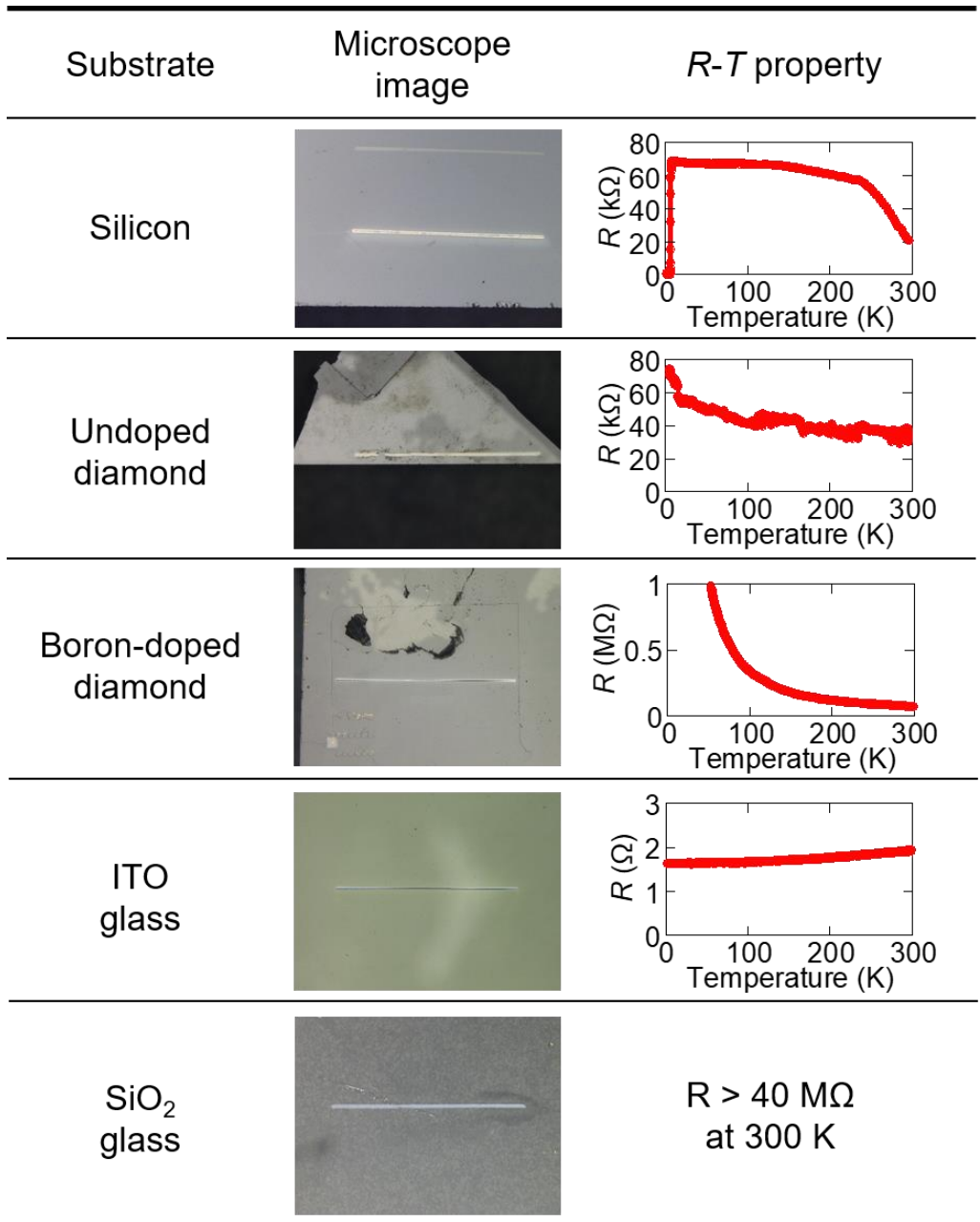

**Figure 3. Optical microscope images and resistance-temperature ($R$-$T$) properties of various gallium-irradiated substrates of silicon, undoped diamond, boron-doped diamond, ITO (indium tin oxide) glass, and SiO$_2$ glass.**

The gallium-irradiated silicon of square pattern was used for chemical state analysis using XPS. Figure 4 (a) shows the depth profile of core-level Ga 2p XPS spectra in the gallium-irradiated region of silicon substrate. The etching treatment was performed by an Ar gas cluster ion beam (GCIB) with 10 keV beam energy. The mean size of one cluster was approximately 1000 atoms, the scanning area of the GCIB was about 2 mm$^2$, and the beam current was about 5 nA. The GCIB mills the sample surface very slowly without a change of the intrinsic chemical state during the irradiation [30]. According to the surface spectrum, we can see the two individual peaks around 1119.1 eV and 1116.6 eV, corresponding to the pure gallium peak [31] and the most stable oxide (Ga$_2$O$_3$) peak [32], respectively. The oxide peak was gradually decreased by the GCIB etching and completely disappeared at the etching depth of 4.8 nm. The literature regarding to the gallium implantation for n-type silicon substrate with 30 nm thick $SiO_2$ reported that the surface spectrum showed no signal of Ga 2p [18]. When the 14 nm depth was milled, the $Ga_2O_3$ peak appeared, and pure gallium peaks was observed from 18 nm depth, according to the previous report of depth profile [18]. Because the acceleration voltage for the gallium implantation is quite lower in our FIB process than 80 keV of the previous study [18], it could be considered that the irradiated gallium stayed at shallow region around surface with high concentration.

Figure 4 (b) shows the core-level Si 2p XPS spectra. The upper spectrum was acquired from the gallium-irradiated region and lower one was from the other region on the same silicon substrate. The Si 2p photoemission is split into two peaks, one at high energy side attributed to Si $2p_{1/2}$ and the lower one attributed to Si $2p_{3/2}$. The shape of Si 2p peak exhibits a broadening feature after the gallium irradiation from a comparison between two spectra. The full width at half maximum (FWHM) of 0.57 eV in the as-prepared silicon was change to 0.68 eV in the gallium-irradiated region. The peak broadening is known as a signal from an amorphization which generated by the ion scattering around the surface [33]. These peak changes indicate that there are various states of Si-Si bonding distance and bonding angle around the surface, and maybe it affects the superconducting properties.

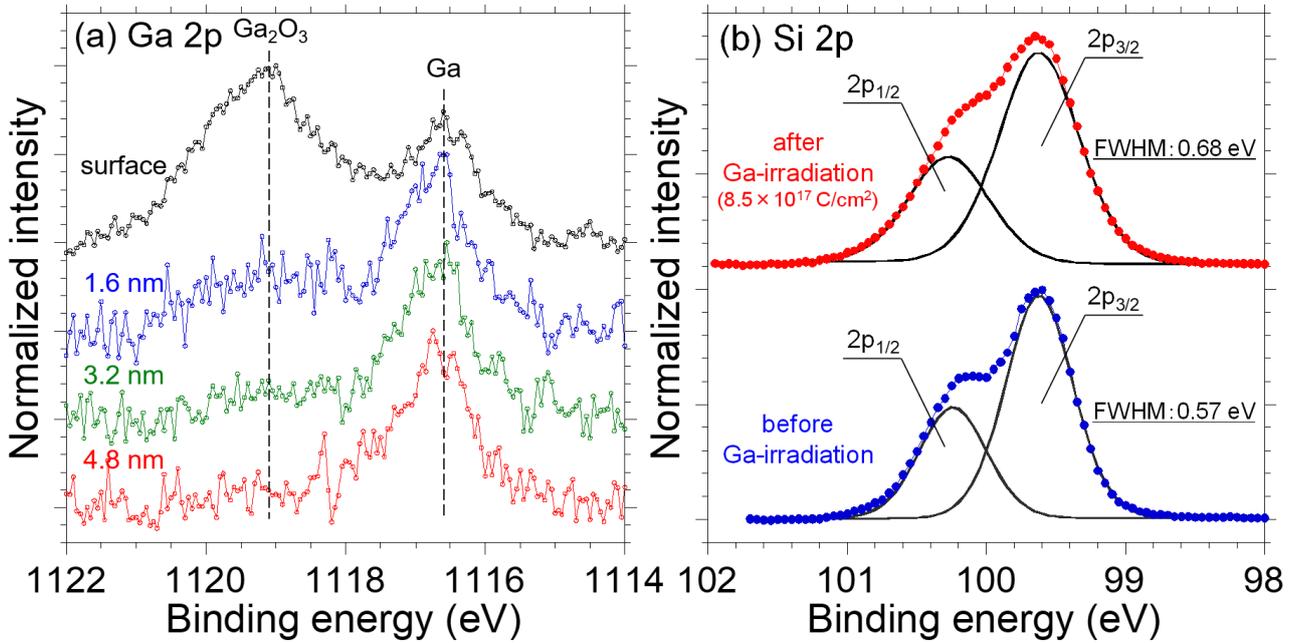

**Figure 4. (a) Depth profile of core-level Ga 2p XPS spectra in the gallium-irradiated region of silicon substrate. (b) Core-level Si 2p XPS spectra in the gallium-irradiated region and as-prepared silicon substrate. The pattern is square (200×200×0.3 μm) with dose amount of $8.5×10^{17}$ C/cm².**

It is important to determine the critical dose amount for the superconductivity in the gallium-irradiated silicon for the practical application. The gallium irradiations with same dimension (200×200 μm square) by dose amounts of $37×10^{15}$ C/cm², $227×10^{15}$ C/cm², and $850×10^{15}$ C/cm², were performed to examine the transport properties. Figure 5 shows the dose amount dependence of the resistance at 300 K in the square pattern of gallium-irradiated silicon. The lowest dose sample showed quite high resistance of $10^5$ Ω order. The resistance dramatically decreased less than $10^3$ Ω order with increase of the dose amount. The decrease of resistance tended to saturate below $10^3$ Ω. It means the gallium implantation rate is first proportional to dose amount. When the dose amount achieves a certain criterion, the implantation rate and etching rate become comparable, and then the gallium implantation saturates. The inset shows the temperature dependence of each gallium-irradiated samples. The resistance in the lowest dose sample ($37\times10^{15}$ C/cm$^2$) drastically increased as a function of temperature, and it showed no sign of superconductivity at least 2 K. On the other hand, although the middle dose sample ($227\times10^{15}$ C/cm$^2$) indicated clear superconducting transition with $T_c^{onset}$ ~5 K, the zero resistance was not observed. The highest dose sample ($850\times10^{15}$ C/cm$^2$) which is same as the fig.2 (e) showed clear zero-resistance. These results indicated that the critical dose amount to obtain the zero-resistance is between $227\times10^{15}$ C/cm$^2$ and $885\times10^{15}$ C/cm$^2$. It is expected that the device fabrication, such a SQUID magnetometer, by using this mask-less patterning technique of superconducting circuit will be highly anticipated.

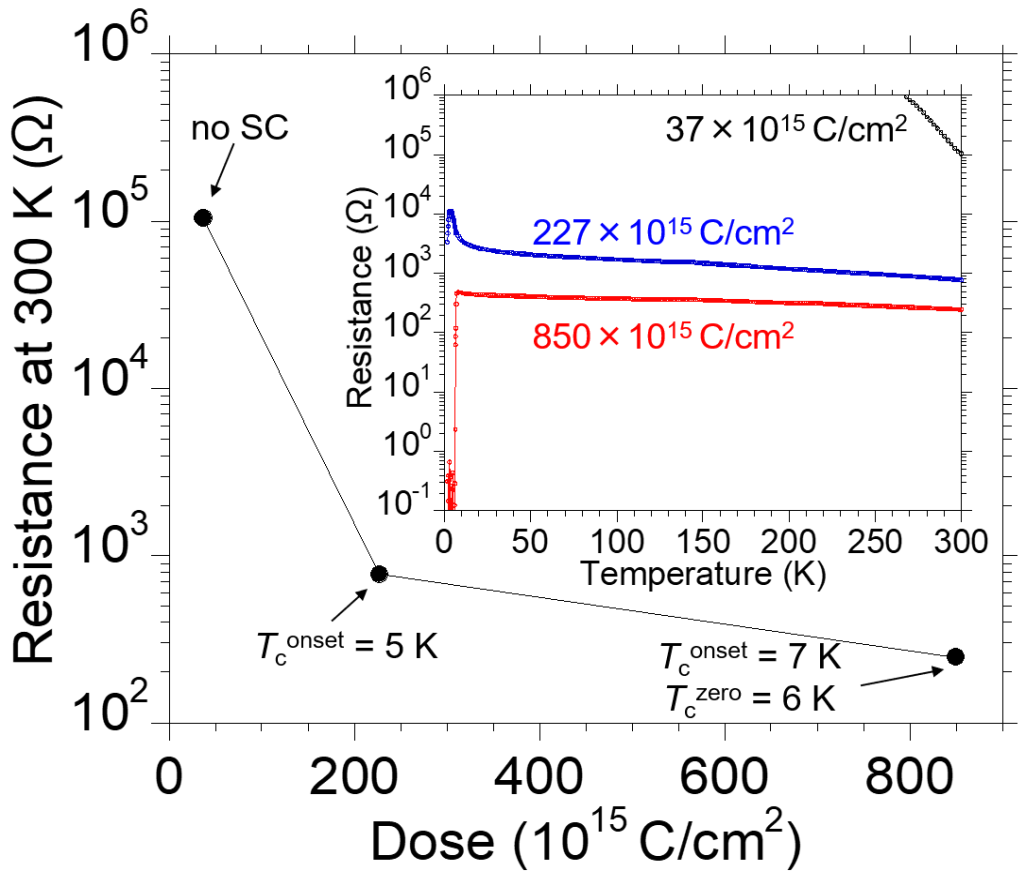

**Figure 5. Dose amount dependence of the resistance at 300 K in the square pattern of gallium-irradiated silicon. The inset is the temperature dependence of each gallium-irradiated samples of the square pattern (200×200×0.3 μm) with different dose amount.**

## 4. Conclusion

In this study, a nanofabrication technique for mask-less patterning of superconducting region by gallium-irradiated silicon on the substrate using FIB was introduced. In various substrates of silicon, diamond, boron-doped diamond, ITO glass, and SiO$_2$ glass, only silicon substrate showed superconductivity with onset $T_c$ of 7 K after gallium irradiation. The line and square shapes of gallium irradiated silicon were fabricated to confirm the versatility of patterning. Although both patterns exhibited superconductivity, very large antistrophic behavior against applied magnetic field was observed in square pattern. In the line pattern, we observed a signature of higher $T_c^{onset}$ above 10 K maybe due to an inhomogeneity of dose amount. The depth profile of XPS spectrum revealed that the irradiated gallium was shallowly distributed in the silicon surface. The surface of irradiated silicon changed to the amorphous like state from the peak broadening of Si 2p XPS spectra. The critical dose amount for superconductivity is between $227 \times 10^{15}$ C/cm$^2$ and $885 \times 10^{15}$ C/cm$^2$. This direct patterning technique of superconducting circuit on silicon substrate without any masks significantly contributes to the application of superconducting devices, such as SQUID magnetometer.

**Acknowledgment**

The authors thank Prof. Dr. T. Yamaguchi and Mr. Sasama (NIMS) for the early experiments, and also thank Mr. J. Aoto (NIT, Yonago College) for the supports regarding to the sample preparations and measurements. This work was partly supported by JST CREST Grant No. JPMJCR16Q6, JST-Mirai Program Grant Number JPMJMI17A2, JSPS KAKENHI Grant Number JP17J05926 and 19H02177.